\documentclass[
    prd, superscriptaddress, nofootinbib, amsmath, amssymb,
    aps, floatfix, preprintnumbers, twocolumn
]{revtex4-2}
\usepackage[utf8]{inputenc}
\usepackage[dvips]{graphicx}
\usepackage[dvipsnames]{xcolor}
\definecolor{CiteBlue}{RGB}{45,52,151}
\usepackage[
    colorlinks=true,
    linkcolor=CiteBlue,
    urlcolor=CiteBlue,
    citecolor=CiteBlue
]{hyperref}
\usepackage{aas_macros}
\usepackage{booktabs}

\usepackage[capitalise]{cleveref}
\usepackage{siunitx}
\DeclareSIUnit{\year}{yr}
\DeclareSIUnit{\au}{AU}
\DeclareSIUnit{\msol}{M_\odot}

\newcommand{\refcite}[1]{Ref.~\cite{#1}}
\newcommand{\refscite}[1]{Refs.~\cite{#1}}

\usepackage{bm}
\newcommand{\bb}[1]{\bm{\mathrm{#1}}}
\newcommand{\du}{\mathrm{d}}
\newcommand{\dd}{\,\du}
\newcommand{\erfc}{\operatorname{erfc}}
\newcommand{\tot}{\mathrm{tot}}
\newcommand{\prob}{\operatorname{Pr}}
\newcommand{\pobs}{p}
\newcommand{\obs}{\mathrm{obs}}
\newcommand{\avgb}{\left\langle{}b\right\rangle}
\newcommand{\many}{\mathrm{many}}
\newcommand{\ann}{\mathrm{ann}}
\newcommand{\orb}{\mathrm{orb}}

\usepackage{dcolumn}
\newcolumntype{C}[1]{>{\centering\arraybackslash}p{#1}}

\newcommand{\ed}[1]{\textcolor{magenta}{#1}}

\begin{document}

\title{Fast and Fewrious:\texorpdfstring{\\}{ }Stochastic binary perturbations from fast compact objects}
\preprint{MIT-CTP/5887}

\author{Badal Bhalla}
\email{badalbhalla@ou.edu}
\affiliation{Homer L. Dodge Department of Physics and Astronomy, University of Oklahoma, Norman, OK 73019, USA}

\author{Benjamin V. Lehmann}
\email{benvlehmann@gmail.com}
\affiliation{Center for Theoretical Physics -- a Leinweber Institute, Massachusetts Institute of Technology, Cambridge, MA 02139, USA}

\author{Kuver Sinha}
\email{kuver.sinha@ou.edu}
\affiliation{Homer L. Dodge Department of Physics and Astronomy, University of Oklahoma, Norman, OK 73019, USA}

\author{Tao Xu}
\email{tao.xu@ou.edu}
\affiliation{Homer L. Dodge Department of Physics and Astronomy, University of Oklahoma, Norman, OK 73019, USA}

\date{\today}

\begin{abstract}
Massive compact objects soften binaries. This process has been used for decades to constrain the population of such objects, particularly as a component of dark matter (DM). The effects of light compact objects, such as those in the unconstrained asteroid-mass range, have generally been neglected. In principle, low-energy perturbers can harden binaries instead of softening them, but the standard lore is that this effect vanishes when the perturber velocities are large compared to the binary's orbital velocity, as is typical for DM constituents. Here, we revisit the computation of the hardening rate induced by light perturbers. We show that although the perturbations average to zero over many encounters, many scenarios of interest for DM constraints are in the regime where the variance cannot be neglected. We show that a few fast-moving perturbers can leave stochastic perturbations in systems that are measured with great precision, and we use this framework to revisit the constraint potential of systems such as binary pulsars and the Solar System. This opens a new class of dynamical probes with potential applications to asteroid-mass DM candidates.
\end{abstract}

\maketitle

\section{Introduction}
\label{sec:introduction}
The search for dark matter (DM) has always spanned an enormous range of candidates and scales. For the past several decades, the field has been dominated by the prospect of discovering weakly interacting massive particles (WIMPs) at $\mathcal{O}(\qty{100}{\giga\electronvolt})$, but a wide variety of other possibilities have been recognized from the earliest days of DM science~\cite{deSwart:2017heh}. Now that collider searches and laboratory experiments have placed the WIMP paradigm under pressure~\cite{Baer:2020kwz,Arcadi:2017kky,Arcadi:2024ukq}, other classes of candidates at much lower and much higher mass scales are resurgent, from axion-like particles at $\mathcal O(\qty{e-20}{\electronvolt})$ to dark compact objects at masses up to \qty{100}{\msol}.

The latter category subsumes many DM candidates predicted by a wide range of microphysical scenarios~\cite{Baryakhtar:2022hbu}. These include Q-balls~\cite{Coleman:1985ki,Kusenko:1997si}, boson stars~\cite{Jetzer:1991jr,Liebling:2012fv}, axion miniclusters~\cite{Hogan:1988mp}, and, most prominently, primordial black holes (PBHs)~\cite{Zeldovich:1967lct,Hawking:1971ei,Carr:1974nx,Carr:2016drx,Green:2020jor,Carr:2020xqk,Green:2024bam,Carr:2023tpt}. Dark compact objects can contribute to the DM density over an enormous span of masses, but the parameter space of greatest interest at present is the so-called asteroid mass range, roughly \num{e-17}--\qty{e-12}{M_\odot} (\num{e17}--\qty{e22}{\gram}). In this window, no observational probes currently rule out compact objects as accounting for all of the DM, despite many ongoing efforts~\cite{Gorton:2024cdm}. While microlensing is effective at higher masses, the typical Einstein radius of an asteroid-mass object is too small, rendering such probes ineffective due to the nonzero angular size of light sources and the nonzero wavelength of the light itself~\cite{Niikura:2017zjd,Sugiyama:2019dgt,Smyth:2019whb}. Additionally, PBHs in particular are unconstrained by evaporation in this regime, meaning that simple PBH models can account for the DM abundance without invoking any new degrees of freedom beyond those required for cosmic inflation~\cite{Clesse:2017bsw}.

Constraining asteroid-mass objects requires qualitatively new probes. One promising direction leverages the same philosophy that has historically been applied to constrain MAssive Compact Halo Objects (MACHOs) at masses of order \qty{100}{\msol} and above: dynamical effects. It has long been recognized that if the DM consists of ultramassive objects, then encounters between these objects and stars will result in large perturbations to stellar trajectories. This was used to develop what may have been the very first constraint on the DM mass: in 1969, \refcite{1969Natur.224..891V} argued that due to tidal distortions, DM could not be composed of objects with masses between \num{e8} and \qty{e13}{\msol}. Since then, with the advent of more sophisticated theoretical tools and much larger datasets, these constraints have been substantially refined, and provide robust limits on compact objects as DM at masses above $\mathop{\sim}\qty{100}{\msol}$. Recently, several groups have proposed the use of high-precision timeseries measurements to search for perturbations induced by much smaller compact objects~\cite{Tran:2023jci,Bertrand:2023zkl,Thoss:2024vae}, with potential sensitivity in the unconstrained asteroid-mass window. Other proposals take advantage of dynamical processes that can leave a light perturber trapped in the system~\cite{Bhalla:2024jbu,Lehmann:2022vdt,DeLorenci:2025wbn}. These ideas rely on the exquisite precision of measurements within the Solar System or in the immediate Solar neighborhood.

However, it is also possible that a wide variety of other systems might provide useful dynamical probes of light compact objects. Here, it is instructive to revisit the mechanisms that have been used in the past to constrain more massive compact objects. The state of the art uses perturbations to wide binary systems~\cite{Penarrubia:2010pa,Penarrubia:2016ltr,Ramirez:2022mys,2009MNRAS.396L..11Q, Brandt:2016aco, Tyler:2022rxi} or dynamical heating of cold systems~\cite{Graham:2023unf}. The latter method provides a particularly intuitive explanation for the efficacy of constraints at this mass scale. The velocity distributions of stars and DM particles are comparable, each being determined by the Galactic potential. If the DM particle is much more massive than stars, then the effective temperature of the DM fluid is higher than that of the stellar fluid, meaning that energy is inevitably transferred as heat from the DM to the stellar distribution.

In the opposite limit, where the DM particle is much lighter than stars, the stars should transfer energy to the DM. This indeed takes place via dynamical friction~\cite{Chandrasekhar:1943ys}. One might thus hope to observe energy loss from binaries (i.e., hardening), as opposed to the energy gain (softening) used as a probe by \refscite{Penarrubia:2010pa,Penarrubia:2016ltr,Ramirez:2022mys,2009MNRAS.396L..11Q, Brandt:2016aco, Tyler:2022rxi}. However, probing DM with dynamical friction is extremely challenging, and no known systems are capable of directly probing the dark matter abundance expected in their environments~\cite{Caputo:2017zqh}. A major difficulty is the hierarchy of velocities between the DM and stellar binary components. As already anticipated in Chandrasekhar's original work~\cite{Chandrasekhar:1943ys}, the dynamical friction induced by particles much faster than the binary's orbit is negligible. Typical binaries have orbital velocities of $\mathcal O(\num{0.1}\textnormal{--}\qty{10}{\kilo\meter/\second})$, whereas the DM velocity in the Galactic halo is $\mathcal O(\qty{230}{\kilo\meter/\second})$. Thus, dynamical friction from DM on stellar binaries is strongly suppressed.

But dynamical friction is not the only means by which pertubers can harden a binary. Three-body processes feature complicated dynamics that can either soften or harden a binary, depending on the parameters of the encounter. As a rule of thumb, Heggie's law~\cite{Heggie:1975rcz} states that binaries that are soft with respect to the perturber (i.e., with binding energy lower in magnitude than the energy of the perturber) are likely to soften further in an encounter, whereas binaries that are hard with respect to the perturber (i.e., with binding energy greater in magnitude) are likely to harden further. Since we are interested in light perturbers, which have accordingly low energy, typical binaries are already hard according to this classification, and thus might be further hardened by repeated encounters with light perturbers. This is a distinct effect from dynamical friction, which is an essentially many-body effect, and scales differently with the parameters of the binary and the perturbers.

Despite the differences between dynamical friction and three-body hardening, one caveat remains: the na\"ive interpretation of Heggie's law breaks down when the perturbers are much faster than the orbital velocity of the binary. \refscite{Gould:1990bk,Quinlan:1996vp} have shown that in this regime, the average energy transferred from the binary to the perturber quickly vanishes. Thus, three-body hardening by dark compact objects is often neglected. However, as we will explain, three-body hardening is only suppressed in this regime when \emph{averaged} over many encounters. If the number density of perturbers is low, a typical observing period may only feature one or several encounters. In this case, while the mean energy transferred is close to zero, the \emph{width} of the distribution of energy transfers is not. This suggests a new opportunity in the regime of \emph{rare, high-velocity} perturbers: the fast and the few.

In this work, we investigate such \emph{stochastic} binary hardening due to light compact objects. We revisit the computation of the hardening rate originally performed by \refscite{Gould:1990bk,Quinlan:1996vp}, and develop a framework for estimating the energy transferred cumulatively over a small number of encounters. Since these perturbations can have either sign, the net effect is not ``hardening,'' per se, but rather a kind of Poisson noise. We thus revive three-body dynamics as a probe of light compact objects, just as three-body softening has proven so valuable as a probe of massive compact objects.

This work is structured as follows. In \cref{sec:three-body}, we review the physics of three-body encounters, and discuss the evolution of Heggie's law from the original proposal in \refcite{Heggie:1975rcz} to the revised versions of \refscite{Gould:1990bk,Quinlan:1996vp}. In \cref{sec:multiple-perturbations}, we update Heggie's law further to correctly account for stochasticity in the limit of few perturbations, and we develop a full estimate of the hardening rate in this regime. In \cref{sec:prospects}, we apply these results to a set of realistic systems, and evaluate the prospects for detecting compact objects by this means. We discuss our results and conclude in \cref{sec:conclusions}.

Throughout this work, we denote quantities associated with the two objects in a binary by subscripts 1 and 2, and those associated with the perturber by a subscript 3. In particular, $v_{12}$ denotes the relative velocity of the two components of the binary; $v_3$ denotes the velocity of the perturber relative to the center of mass of the binary; $m_{12} \equiv m_1 + m_2$ denotes the total mass of the binary; and $E_{12}$ denotes the (negative) total mechanical energy of the binary. We denote the semimajor axis of the binary by $a$. We use a superscript star (e.g., $a^\star$) to denote parameter values used in reference simulations.

\section{Three-body scattering}
\label{sec:three-body}
Three-body scattering is notoriously complicated, and outcomes are highly sensitive to initial conditions. Nonetheless, three-body encounters are important in a number of astrophysical contexts, particularly in cases where many encounters cumulatively affect the evolution of a binary. As a result, various heuristic and statistical approaches have been developed in the astrophysical literature to classify and predict the outcomes of an ensemble of three-body encounters. We now review the features of three-body encounters that have been well studied, and we situate the current work in that context.

\subsection{Outcomes of three-body encounters}
\begin{figure}[h]
\centering
\includegraphics[width=1\linewidth]{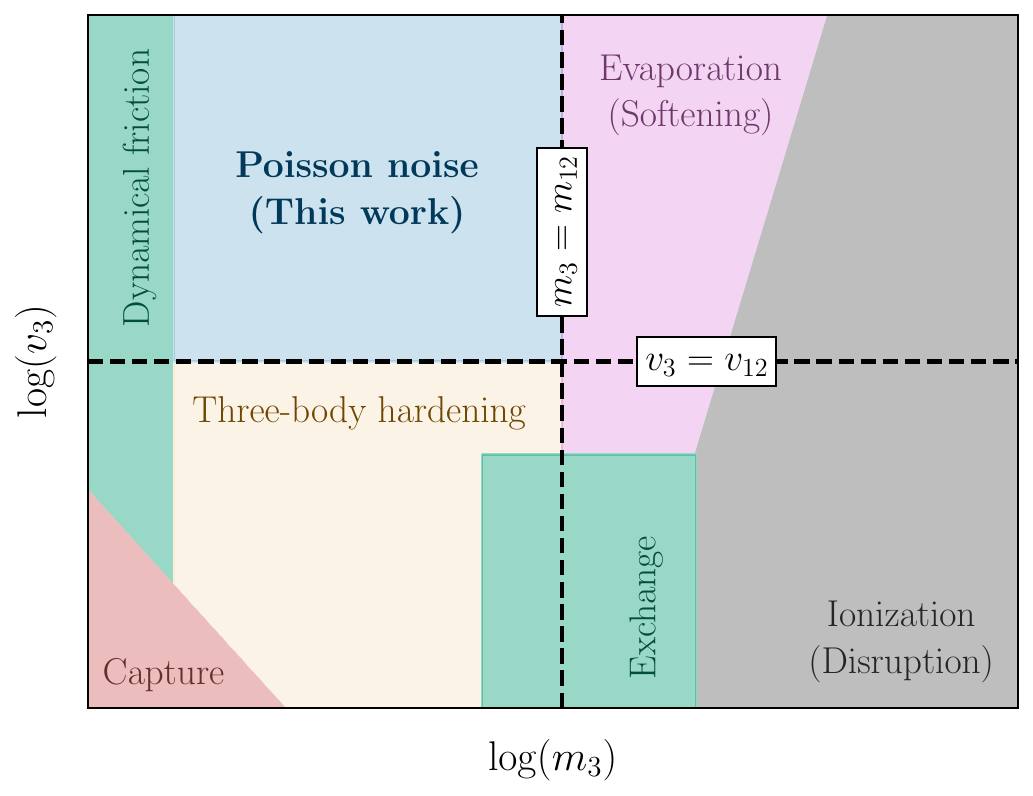}
\caption{Schematic description of the regimes in which different processes are relevant. These regimes are shown in the plane of perturber mass and velocity, on logarithmic axes. The boundaries between regions are only illustrative: realistically, several of these processes can occur in overlapping regions. Note that ``evaporation'' and ``ionization'' are terms used in the three-body scattering literature to refer to softening and disruption, respectively, and are unrelated to Hawking evaporation and atomic ionization. The dashed black lines indicate the mass and velocity of the binary. In this work, we study the light blue region, where velocities are high (\emph{fast}), but number densities are still low (\emph{few}) compared to the dynamical friction regime.}
\label{fig:schematic}
\end{figure}
Three-body encounters can be usefully classified on an energetic basis. The initial mechanical energy of the binary and the perturber determine important features of the encounter, and outcomes can be classified by the final energies of the objects. Specifically, given a particular perturber, the binary can be categorized as ``hard'' (tightly bound) or ``soft'' (weakly bound) with respect to the perturber, which we will soon define concretely. A classic paper by Heggie~\cite{Heggie:1975rcz} shows that \textit{on average,} a hard binary becomes harder following a three-body encounter, whereas a soft binary becomes softer. This is known as Heggie's law.

This is easy to understand in the limit of an extremely hard or extremely soft binary. Consider a binary system with component masses $m_1$ and $m_2$ and semimajor axis $a$, and a perturber with mass $m_3$ and velocity $v_3$. Now, suppose that $m_3 \gg m_1, m_2$, and suppose that $v_3$ is comparable to the orbital velocity of the binary. During the encounter, the potential binding the components of the binary is small compared to the potential sourced by the perturber, so the interaction of the perturber with each component is well approximated by elastic $2 \to 2$ scattering. In the limit of large $m_3$, corresponding to a very soft binary, the typical energy transfer to each component is large compared to the binding energy. Thus, a generic encounter leads to a final state in which the components are free, i.e., the binary is no longer bound. This is called ``disruption'' or ``ionization,'' and corresponds to extreme softening. On the other hand, for an extremely hard binary with $m_1, m_2 \gg m_3$ and orbital velocity $v_{12} \gg v_3$, the scattering process tends to accelerate the perturber, causing the binary to become more tightly bound (harder).

Disruption is only kinematically allowed for soft binaries. The total energy of the three-body system, measured in the rest frame of the center of mass of the binary, is given by
\begin{equation}
    E_{123} = \frac{1}{2}m_3v_3^2 - \frac{Gm_1m_2}{2a}.
\end{equation}
Disruption entails that all three objects are free in the final state, so the total energy must be positive in the initial state. Thus, conservation of energy only allows for disruption if the binary is sufficiently soft in the initial state. On the other hand, if the total energy of the three-body system is negative, at least two objects must remain bound in the final state, but these need not be the original components of the binary. In certain cases, the perturber can transfer enough energy to one component to disrupt the binary, but lose so much energy in the process that the perturber becomes bound to the other component. These are exchange processes. On the other hand, if the initial energy of the perturber is too low, then it can become bound without unbinding either component of the binary, forming a triple system. These are capture processes.

All of these processes have been used as probes of dark compact objects. The disruption process has been used extensively as a probe of compact objects that are much more massive than stars, i.e., in the regime where binaries are typically soft~\cite{1985ApJ...290...15B,Quinn:2009zg,Allen:2014tla,Penarrubia:2010pa,Monroy-Rodriguez:2014ula,Penarrubia:2016ltr,Ramirez:2022mys,2009MNRAS.396L..11Q, Brandt:2016aco, Tyler:2022rxi}. Capture and exchange processes have also been studied for this purpose~\cite{Lehmann:2020yxb,Lehmann:2022vdt,Bhalla:2024jbu}, as they become relevant for dark compact objects at different masses. However, designing probes based on softening and hardening, beyond the simplest disruption processes, requires a more precise formulation of Heggie's law. The rate of softening and hardening---and indeed, whether these take place at all---cannot be determined directly from the energetics of the perturber. A precise definition of hard and soft binaries, as well as their distinguishing characteristics, is required.

The definition of hard and soft binaries is a topic of significant debate. According to Heggie~\cite{Heggie:1975rcz}, a binary system should be considered hard if its binding energy is much greater than the kinetic energy of the perturber, or soft if the binding energy is much less than the kinetic energy of the perturber. A second definition was given by Hills~\cite{1990AJ.....99..979H}, who found that in numerical experiments, perturbers traveling faster than the binary's orbital speed, $v_{12}$, tend to soften the binary, while those traveling slower tend to harden it. The analytical estimates provided by Heggie align with the numerical experiments conducted by Hills when all three objects involved in the encounter have the same mass. However, if the third object is fast but is significantly less massive than the two binary stars, the two definitions are in conflict.

This is particularly problematic for constraining compact objects as a DM candidate. The objects of greatest interest are those in the asteroid-mass range, much lighter than the components of most observationally-accessible binaries, but DM particles or compact objects have velocities on the order of the dispersion of the Galactic halo, much faster than typical binary orbits. The various outcomes of three- or many-body encounters are illustrated schematically in the plane of perturber mass and velocity in \cref{fig:schematic}, for particular choices of binary parameters. The remaining window for compact object DM is corresponds to the top-left quadrant of the figure (light blue), in the regime where the classical hardening and softening behaviors are not readily applicable. Understanding hardening and softening in this regime requires different methods.

\subsection{Hardening via multiple encounters}
The issue of fast low-mass objects was first addressed by Gould~\cite{Gould:1990bk} and later investigated numerically by Quinlan~\cite{Quinlan:1996vp}. (See also \refcite{2021MNRAS.508..190G}.) We first give a concrete definition of the hardening rate considered in that work. Consider a binary in a background of perturbers with fixed abundance and velocity. \refcite{Quinlan:1996vp} defines a dimensionless energy transfer
\begin{equation}
    C = \frac{m_{12}}{2m_3}\frac{\Delta E_{12}}{E_{12}},
\end{equation}
and the hardening rate is the average value of $C$ over all of the parameters of a three-body encounter. The hardening rate determines how fast the orbit of the binary with semimajor axis $a$ would shrink in a medium of perturbers with number density $n$, and is defined as

\begin{equation}
    \frac{\du}{\du t}\left( \frac{1}{a} \right) = \frac{G n m_3}{v_3}H_1.
\end{equation}

In practice, we define $\langle C\rangle|_b$ to be the average of $C$ over all angular parameters at fixed impact parameter $b$. \refcite{Quinlan:1996vp} then computes the hardening rate $H_1$ by
\begin{equation}
    \label{eq:hardening-rate}
    H_1 = 8 \pi \int_0^\infty \du b \, \frac{b}{b_0^2} \, \langle C\rangle|_b
    ,
\end{equation}
where $b_0$ is the impact parameter with pericenter distance equal to the semimajor axis, i.e., $r_p = a$, with $r_p$ defined by the relation
\begin{equation}
    \label{eq:impact-parameter}
    b^2 = r_p^2 \left( 1 + \frac{2Gm_{12}}{r_p v_3^2}  \right)
    .
\end{equation}
That is, $b_0^2 = a^2[1 + 2Gm_{12}/(av_3^2)]$. The definition of $H_1$ averages the dimensionless energy transfer over possible impact parameters, weighted by their cross sections: the cross section for an impact parameter $b$ corresponds to the area of a circle of radius $b$, so the appropriate measure of integration is $b\dd b$. The definition of $b_0$ accounts for gravitational focusing: slow objects have pericenter distances much smaller than their impact parameters. 

Notice that the hardening effect from multiple successive three-body encounters is distinct from the traditional definition of dynamical friction. Dynamical friction slows a massive object at a rate
\begin{equation}
    \dot{\bb v}_1 = -\frac{16\pi^2\log(\Lambda)G^2m_1\rho_3\bb v_1}{v_3^3}
        \int_0^{v_1}\du v\,v^2 f(v)
    ,
\end{equation}
where $f$ is the velocity distribution function and $\log\Lambda$ is a Coulomb logarithm, typically $\mathcal O(10)$. This is second-order in $G$. Physically, this is because dynamical friction is due to the interaction of a massive object with its wake in the density field: as particles pass by the massive object, they are gravitationally focused into an overdensity on the other side, which then pulls the massive object back. Dynamical friction couples the velocity of a single massive object to the average velocity of the medium, independent of the three-body dynamics that take place when perturbers interact with a binary. Thus, although there are some similarities between dynamical friction and three-body hardening, the two are not the same, and the effects we discuss here need not match onto dynamical friction in the low-mass limit.

\begin{figure*}
    \includegraphics[width=\linewidth]{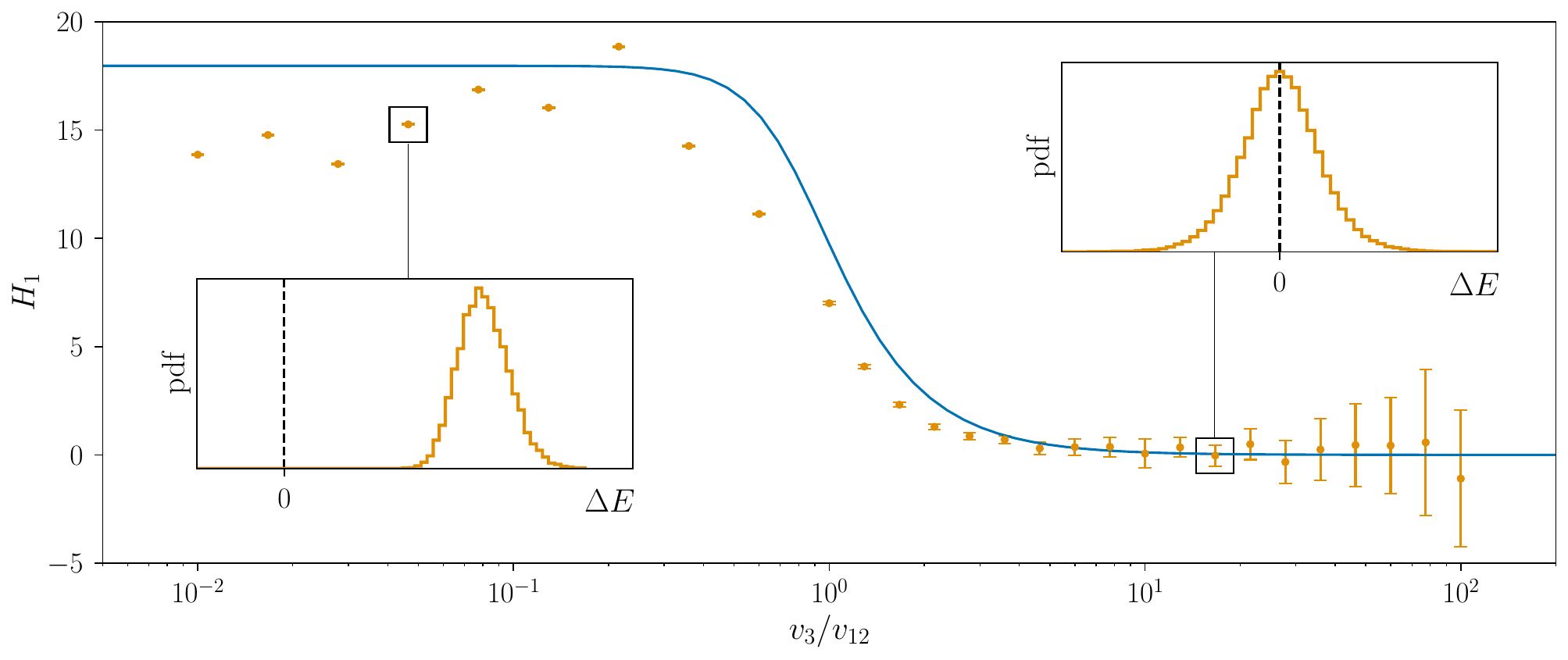}
    \caption{Hardening rate as a function of velocity. Orange points are computed from an ensemble of simulations. The blue curve shows the fit from \refcite{Quinlan:1996vp}. Error bars indicate $1\sigma$ uncertainties on the mean values. Insets show the full distribution of energy transferred in a single encounter for selected velocities, and are shown to scale. For high perturber velocities $v_3 \gg v_{12}$, while the mean energy transfer goes to zero, the width of the distribution remains nonzero.}
    \label{fig:hardening-velocity}
\end{figure*}

\Cref{fig:hardening-velocity} shows the hardening rate as a function of perturber velocity, as computed by Quinlan~\cite{Quinlan:1996vp}. Quinlan further provides a fitting formula for the velocity dependence, as 
\begin{equation}
    H_1(v_3) = H_1(0)\left[
        1 + \left(\frac{v_3}{v_{12}}\right)^4
    \right]^{-1/2}
    ,
\end{equation}
from which we may note that the hardening rate is suppressed by $(v_3 / v_{12})^{-2}$ for large perturber velocities $v_3 \gg v_{12}$. Based on these findings, Quinlan proposed a new convention for classifying hard binaries: a binary system with masses $m_1 \geq m_2$ should only be considered hard if the perturber velocity falls below a critical threshold of the form
\begin{equation}
    v_3 \lesssim 0.85\,v_{12} \sqrt{\frac{m_2}{m_{12}}}
    ,
\end{equation}
where the value of the prefactor and the scaling behaviors with masses were extracted from numerical experiments.

To understand this definition, observe that further hardening of the binary in a three-body encounter corresponds to acceleration of the perturber by the gravitational slingshot mechanism. Here, the perturber ($m_3$) has a close encounter with the lighter of the two bodies ($m_2$), undergoing a process that is very nearly elastic two-body scattering in the center-of-mass frame of $m_2$ and $m_3$. Since $m_2$ and $m_3$ have a close encounter, each mass experiences the same acceleration due to $m_1$, so the encounter can be treated as a two-body scattering process that takes place in an accelerated reference frame. Thus, in the center-of-mass frame of $m_2$ and $m_3$, the speed of recession is equal to the speed of approach, so the encounter corresponds to a deflection of the perturber. When this final state is transformed out of the accelerated $m_2$-$m_3$ frame, back to the inertial center-of-mass frame of the entire three-body system, this deflection corresponds to the acceleration or deceleration of the perturber, depending on the details of the encounter.

If $m_3 \ll m_2 \ll m_1$, then the center-of-mass frame of $m_2$ and $m_3$ is approximately the rest frame of $m_2$ alone, and the center-of-mass frame of the entire system is approximately the rest frame of $m_1$, which sits at the barycenter. The accelerated $m_2$-$m_3$ frame follows the orbit of $m_2$ around $m_1$. In this case, working in the barycentric rest frame, where $v_1 \approx 0$, the mass $m_3$ can only be accelerated (i.e., the binary can only be hardened) if $v_3 < v_2$ prior to the encounter. This means that some hardening processes are only available for low-velocity perturbers, even at very low energies. The critical perturber velocity below which the binary should be considered hard cannot be inferred strictly from the relative energy of the binary and the perturber: the relative velocities themselves are necessary, regardless of the masses. 

The dynamics involved in hardening are quite complicated, so, despite these arguments, the speed at which the transition from hard to soft occurs is difficult to identify from first principles. Quinlan's seminal work was oriented towards computing the hardening rate of massive black hole binaries. Such massive binaries have high velocities, typically larger than the Galactic dispersion velocity. As such, it was not necessary for Quinlan to evaluate the hardening rate for $v_3 \gg v_{12}$, nor would this have been particularly interesting: the evolution of massive black hole binaries proceeds over long timescales subsuming many encounters, so the average hardening rate can be used directly, and the $(v_3/v_{12})^{-2}$ suppression then implies that hardening due to fast-moving perturbers is negligible.

However, when considering encounters between a MACHO and a system such as a binary pulsar, the situation is very different. Firstly, the binary velocity is much lower, so perturbers with the Galactic dispersion velocity are quite fast: $v_3 \gg v_{12}$. Secondly, the period of observation is much shorter, with far higher precision. Measurements of binary pulsars in particular are exquisitely sensitive to the gain or loss of energy, but unlike massive black hole binaries, the evolution of the system is not influenced by many successive encounters over the relevant timescale. For some compact object populations, the expected number of close encounters in the observing period is $\mathcal O(1)$. This means that while the \textit{average} hardening rate does suffer from a substantial suppression, the average is not necessarily the appropriate quantity.

In particular, while the average value of the hardening rate is quickly suppressed at large $v_3$, the rms value of the energy transfer decays much more slowly. As shown in the insets of \cref{fig:hardening-velocity}, the width of the distribution of energy transfers across encounters is nearly the same for perturbers in the slow and fast regimes, and this width is not much smaller than the mean value itself for slow perturbers. This suggests that the most promising route for detecting encounters of this type might lie in their \textit{stochasticity:} while many encounters average away to null effect, the Poisson noise associated with a few encounters might be detectable. We thus label the top-left quadrant of \cref{fig:schematic} as the ``shot noise'' regime, and in the next section, we study the statistics of these stochastic perturbations in detail.

\section{Statistics of multiple perturbations}
\label{sec:multiple-perturbations}
As summarized in the preceding section, it is well known that fast-moving perturbers will neither harden nor soften a binary, on average. That is, the distribution of perturbations delivered to the energy of a binary has a mean of zero. However, this distribution still has nonvanishing width, and if there are not very many encounters over a given period of measurement, there is a nonvanishing stochastic hardening rate expected. In this section, we develop a set of semianalytical estimates for the size of these perturbations, calibrate them using simulations of three-body encounters, and validate them using Monte Carlo sampling.

\subsection{Analytical framework}
To understand the origin (and eventual disappearance) of these stochastic fluctuations, suppose that each pertuber encounter imparts a change $\delta E$ in the energy of the binary. The perturbation $\delta E$ is a random variable whose distribution has some variance $\sigma_E^2$. Now, consider several successive perturbations $\delta E_1,\dotsc,\delta E_N$ over the observing period. By the central limit theorem, for large $N$, the total perturbation $\Delta E^{(N)} = \delta E_1 + \dotsb + \delta E_N$ is normally distributed with a variance of $N\sigma_E^2$. For light perturbers with $m_3 < m_{12}$, observe that $\delta E$ scales linearly with the mass of the perturber, meaning that $\sigma_E^2 \propto m_3^2$, while $N \propto 1/m_3$ if the mass density of perturbers is held fixed. This means that $N\sigma_E^2 \propto m_3$, so at large $N$, $\Delta E^{(N)}$ is normally distributed with a standard deviation that scales as $m_3^{1/2}$, vanishing in the low-mass limit.

In other words, stochastic perturbations arise because the energy transferred in each encounter has a high variance but a mean of zero. The total energy transfer thus behaves as a random walk, with an expectation scaling with the square root of the number of encounters. The number of encounters is inversely proportional to the mass, but this growth at low mass is compensated by the linear mass dependence of the energy transfer in each encounter.

In the rest of this subsection, we make this picture firmly quantitative. Firstly, we define a figure of merit for comparison with observations. We then identify the scaling behavior of this quantity with the parameters of the perturbations. Finally, we define an estimator that combines the effects of many perturbations to serve as a quick indicator of observational accessibility.

\subsubsection{From encounters to detection probabilities}
For observational purposes, the most important quantity is not the expectation value of the total energy transfer, but rather, the probability that the total energy transfer exceeds some threshold $E_{\min}$. We denote this probability by $\pobs_{\obs}$. Formally, $\pobs_{\obs}$ can be written as
\begin{equation}
    \label{eq:p-obs}
    \pobs_{\obs} =
    \sum_{k=1}^\infty \prob(N=k)\times \prob\Bigl(
        \bigl|\Delta E^{(k)}\bigr| > E_{\min}
    \Bigr)
    ,
\end{equation}
i.e., the probability of exceeding the threshold with exactly $k$ encounters weighted by the probability of having exactly $k$ encounters, summed over all values of $k$.

The probability $\prob(N=k)$ of encountering exactly $k$ perturbers is just the pmf of the Poisson distribution $\operatorname{Poiss}(\lambda)$, where $\lambda$ is the average number of perturbers expected to interact with the system over the observing period. Computing the exact distribution of $\Delta E^{(k)}$ is challenging for a general distribution of $\delta E$, but we can easily approximate $\pobs_{\obs}$ in two regimes. First, if $\lambda \ll 1$, we can neglect all terms $k > 1$, so that we need only $\Delta E^{(1)}$, which has the same distribution as $\delta E$. On the other hand, if $\lambda \gg 1$, we can approximate the Poisson distribution as a delta distribution at its mode, writing $\pobs_{\obs} \approx \prob\bigl(\bigl|\Delta E^{(\lambda)}\bigr| > E_{\min}\bigr)$. We can then use the central limit theorem to evaluate the distribution of $\Delta E^{(\lambda)}$, obtaining
\begin{equation}
    \label{eq:p-obs-regimes}
    \pobs_{\obs}(\lambda) \approx
    \begin{cases}
        \displaystyle
        \lambda e^{-\lambda} \prob\bigl(\left|\delta E\right| > E_{\min}\bigr)
        & \lambda \ll 1, \\[3mm]\displaystyle
        \erfc\left(
            \frac{E_{\min}}{\sigma_E\sqrt{2\lambda}}\right)
        & \lambda \gg 1.
    \end{cases}
\end{equation}

The quantity $E_{\min}/(\sigma_E\sqrt{2\lambda})$ scales with $m_3^{-1/2}$, so at large $\lambda$ (small $m_3$), the probability $\pobs_{\obs}$ is exponentially suppressed. However, when $\lambda$ is $\mathcal O(1)$, $\pobs_{\obs}$ can be significantly enhanced over the case of a normal distribution. Specifically, $\pobs_{\obs}$ is enhanced for rare encounters if the distribution of $\delta E$ has heavier tails than a normal distribution with the same variance. We will see that this is indeed the case for three-body encounters.

\subsubsection{Scalings with encounter parameters}
We can also estimate the dependences of $\pobs_{\obs}$ on other parameters of the perturbers. In particular, as long as the encounter is fast, we can use the impulse approximation. Assuming that the imparted momentum to the binary, $\delta p_{12}$, is small compared with the momentum $p_i$ of each component, the imparted energy is given by $\delta E \simeq (p_{12}/m_{12})\,\delta p_{12}$. The impulse approximation predicts that the imparted momentum scales as $\delta p_{12} \propto m_3/v_3$, and therefore, so does $\delta E$.

We can also predict the behavior with impact parameter, at least at large $b$. Here, while the impulse approximation would predict a $b^{-1}$ scaling for the momentum imparted to a single component, the perturbation to binary parameters depends only on the difference between the momenta imparted to the two components: if $m_1$ and $m_2$ are given the same impulse, the only perturbation is to the center of mass of the binary, not to the relative configuration of the components. When the perturber encounters one component with impact parameter $b$, the impact parameter with respect to the other differs by $\mathcal O(a)$. Thus, while the momentum imparted to the first component scales with $b^{-1}$, the momentum imparted to the second is of order $(b+a)^{-1}$, so the difference between them scales with $b^{-1} - (b+a)^{-1}$, or $a/b^2$ for large $b$. This is equivalent to the statement that the binary is perturbed by the tidal potential, for which the impulse goes like $1/b^2$ for $b \gg a$. For $b < a$, an exact prediction is difficult, but the dependence on $b$ should be much gentler, and we approximate it as constant, imposing continuity at $b=a$. At this point, this is purely an assumption, but we will soon see from numerical simulations that it provides a good fit to the energy transfer as a function of impact parameter. Putting these components together, we have
\begin{multline}
    \delta E(b) \simeq
    \delta E^{\star}
    \left(\frac{m_3}{m_3^{\star}}\frac{m_{12}}{m_{12}^{\star}}\right)
    \left(\frac{v_3}{v_3^{\star}}\frac{a}{a^{\star}}\right)^{-1}
    T(a, b),
    \\
    \label{eq:energy-scaling}
    T(a, b) =
    \begin{cases}
        1 & b < a, \\
        \left(\frac ba\right)^{-2} & b > a,
    \end{cases}
\end{multline}
where $\delta E^\star$ is the energy transfer for reference parameters $m_3^{\star}$ and $v_3^\star$, which must be determined by numerical simulations.

\subsubsection{Combining multiple perturbations}
Now, recognizing that $\sigma_E$ has the same scalings as $\delta E$, we can use the large-$\lambda$ limit of \cref{eq:p-obs} to estimate the parametric dependence of the overall probability of observation over many encounters. Consider successive annuli of impact parameters with boundaries $\{b_1,\dotsc,b_n\}$. The contribution of each annulus to the total perturbation $\Delta E$ is a normally distributed random variable with standard deviation $\sigma_k = \sqrt{\lambda_k}\sigma_E\times[\delta E(b_k)/\delta E^{\star}]$, where $\lambda_k$ is the average number of objects passing through the $k$th annulus. The total perturbation is then a normally distributed random variable with variance
\begin{equation}
    \label{eq:sigma-tot}
    \sigma_\tot^2
    = \sum_k\sigma_k^2
    = \sigma_E^2 \left(
        \frac{m_3 m_{12} v_3^{\star} a^{\star}}
            {m_3^{\star} m_{12}^{\star} v_3 a}\right)^2
        \sum_k\lambda_k T(a, b)
    ,
\end{equation}
and the total probability of an observable perturbation is given by $\pobs_{\obs}\approx\pobs_{\many} \equiv \erfc(E_{\min}/\sqrt2\sigma_\tot)$. The mean number of objects passing through each annulus is given by
\begin{equation}
    \label{eq:lambda-k}
    \lambda_k = \frac{\rho_3v_3\,\Delta t}{m_3}\times \pi(b_{k+1}^2 - b_k^2)
    \simeq \frac{2\pi b_k\,\rho_3v_3\,\Delta t}{m_3}\,\Delta b
    ,
\end{equation}
meaning that the sum in \cref{eq:sigma-tot} can be approximated as an integral in the continuum limit. Strictly speaking, this is doubly approximate, since $\lambda_k \to 0$ as $\Delta b\to 0$, meaning that there is only a small number of encounters. This invalidates the application of the central limit theorem, which only holds for a sum of many random variables. As such, the use of \cref{eq:sigma-tot} in this regime cannot be justified a priori, and constitutes a separate assumption. Notwithstanding this caveat, we have
\begin{align}
    \sum_k\lambda_k T(a, b)
    &\approx
    \int_{b_{\min}}^{b_{\max}}\du b\,\frac{2\pi b\rho_3v_3\,\Delta t}{m_3}
    \,T(a, b)
    \\
    &= \frac{\pi\rho_3v_3\,\Delta t}{m_3}\biggl[
        \left(a^2 - b_{\min}^2\right)\Theta(a - b_{\min})
        \\\nonumber &\qquad\qquad\qquad
        + 2a^2\log\left(\frac{b_{\max}}{\max\{b_{\min}, a\}}\right)
    \biggr]
    .
\end{align}
The $b_{\max}$ dependence persists in the form of a Coulomb logarithm, typically of order 10. However, since this computation only applies for fast encounters, we set $b_{\max}$ such that the crossing time of a perturber fits within the observing period, i.e., $b_{\max} = v_3\,\Delta t$, which can result in a much smaller (or vanishing) Coulomb logarithm. We choose $b_{\min}$ such that $\lambda = \lambda_{\min}$ for $b < b_{\min}$, i.e., $b_{\min} = \sqrt{\lambda_{\min}m_3/(\pi\rho_3v_3\,\Delta t)}$, and we take $\lambda_{\min} = 10$ in computations. Now we can write the probability of an observable perturbation as
\begin{equation}
    \pobs_{\many} \simeq \erfc\left[
        \tfrac{E_{\min}m_3^{\star}m_{12}^{\star}a/(
            \sigma_E v_3^{\star} m_{12} a^{\star}
        )\times
            \sqrt{v_3/(2\pi\rho_3m_3\,\Delta t)}}
        {
            \sqrt{
                \left(a^2 - b_{\min}^2\right)\Theta(a - b_{\min})
                + 2a^2\log\left(\frac{b_{\max}}{\max\{b_{\min}, a\}}\right)
            }
        }
    \right]
    .
\end{equation}
This result indicates that we can neglect the slow-moving tail of the DM velocity distribution: the pdf of the speed distribution scales as $v_3^2$ at small $v_3$, so the factor $\sqrt{v_3/\rho_3}$ in the numerator goes as $1/\sqrt{v_3}$, meaning that the argument of $\erfc$ diverges in the low-velocity limit.

For $\lambda < 1$, the central limit theorem does not apply, and combining perturbations from different impact parameters is nontrivial. However, we can still make a similar estimate using the fact that $\lambda e^{-\lambda}$ is fairly sharply peaked at $\lambda = 1$, with a width of $\mathcal O(1)$. We thus conservatively take just a single annulus $b_{\min} < b < b_{\max}$ such that the width $\Delta b \equiv b_2 - b_1$ is half of the average value of $b$ on the annulus and such that $\lambda = 1$. We neglect all energy transfers originating from encounters with impact parameters outside this annulus. This corresponds to
\begin{equation}
    b_1 = \sqrt{\frac{8m_3}{(9+\sqrt{33})\pi\rho_3v_3\,\Delta t}},
    \quad
    b_2 = \frac14\left(1 + \sqrt{33}\right)b_{\min},
\end{equation}
with $\avgb \approx 0.57\sqrt{m_3/(\rho_3v_3\,\Delta t)}$. However, we must also impose the requirement that $\avgb < b_{\max}$, so we take $b = \min\{\avgb, b_{\max}\}$ and $\Delta b = \frac12b$, and compute $\lambda$ as in \cref{eq:lambda-k}. Then the probability of an observable perturbation from an encounter within just this annulus is
\begin{multline}
    \pobs_{\obs} \approx \pobs_{\ann} \equiv
    \lambda(b) e^{-\lambda(b)}
    \prob\Bigl(\bigl|\delta E(b)\bigr|
    > E_{\min}\Bigr)
    ,\\
    \quad
    b = \min\{\avgb, b_{\max}\}
    .
\end{multline}
An independent conservative estimate is to admit only $0 < b < a$, i.e., impact parameters that cross within the orbit of the binary. This gives a lower bound, and since there is no $b$ dependence in this regime, the estimate is straightforward:
\begin{multline}
    \pobs_{\obs} \approx \pobs_{\orb} \equiv
    \frac{\pi\rho_3v_3b^2\,\Delta t}{m_3}
        \exp\left(-\frac{\pi\rho_3v_3b^2\,\Delta t}{m_3}\right)
    \\
    \times \prob\bigl(\left|\delta E(b)\right| > E_{\min}\bigr),
    \quad
    b = \min\{a, b_{\max}\}
    .
\end{multline}

Since these are two different conservative estimates, the actual probability should be greater than each one. Thus, we conservatively estimate the total probability of an observable perturbation as the maximum of the probability through each of these channels, i.e.,
\begin{equation}
    \label{eq:p-tot}
    \pobs_\tot \approx
    \max\left\{
        \pobs_{\orb},
        \pobs_{\ann},
        \pobs_{\many}
    \right\}
    .
\end{equation}

\subsubsection{An estimator for probability of observation}
In typical scenarios, $\pobs_\tot$ exhibits two peaks as a function of $m_3$, since $\pobs_{\orb}$ and $\pobs_{\ann}$ dominate over $\pobs_{\many}$ but peak at different masses, which we denote by $\hat m_\orb$ and $\hat m_\ann$, respectively. This doubly-peaked behavior is an artifact of the division between these different channels. We thus define an alternative estimator $\hat\pobs$ as
\begin{equation}
    \label{eq:p-hat}
    \hat\pobs(m_3) \equiv
    \begin{cases}
        \min\{\max \pobs_{\orb},\,\max \pobs_{\ann}\}
            & m_3\in[\hat m_\orb, \hat m_\ann] \\
        \pobs_\tot(m_3)
            & \textnormal{otherwise.}
    \end{cases}
\end{equation}
The estimator $\hat\pobs$ will be crucial to our results in later sections, and we will validate it with Monte Carlo sampling shortly.

With this formalism, we can now estimate the probability of an observable perturbation in a wide range of scenarios with only one numerical input: the standard deviation $\sigma_E$ for one set of reference parameters. In the next section, we compute $\sigma_E$ directly using numerical simulations.

\subsection{Calibration with three-body simulations}
\begin{figure*}
    \includegraphics[width=0.48\linewidth]{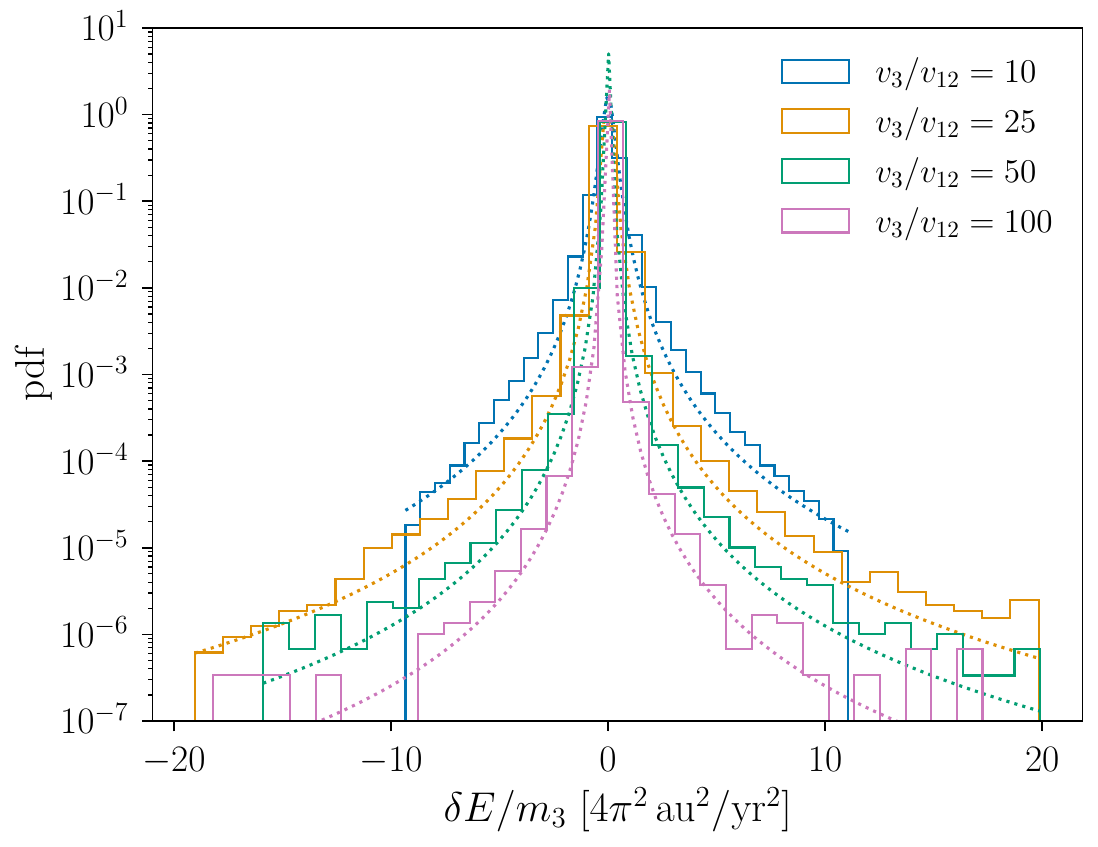}
    \hfill
    \includegraphics[width=0.48\linewidth]{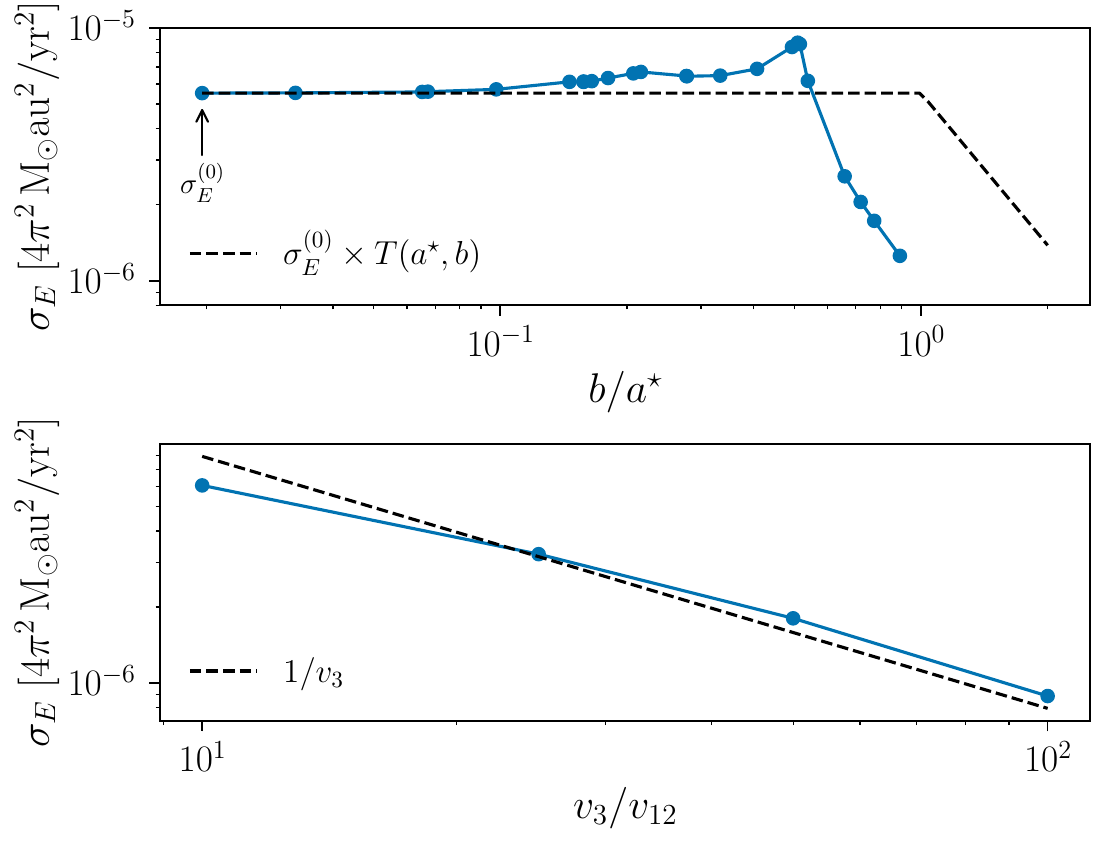}
    \caption{\textbf{Fits to simulation ensembles.} \textit{Left:} Energy transfer distributions in an ensemble of simulations for several values of $v_3/v_{12}$. Dotted curves show the fitted $t$ distribution (see text). \textit{Top~right:} dependence of the variance $\sigma_E$ on the impact parameter of the encounter. The dashed curve shows the behavior implied by our approximation $T(a, b)$ (\cref{eq:energy-scaling}). \textit{Bottom~right:} dependence of $\sigma_E$ on the perturber velocity. The dashed line shows a fit to the $1/v_3$ scaling assumed in \cref{eq:energy-scaling}.}
    \label{fig:fits-sigmas}
\end{figure*}
To directly compute the distribution of the energy transfers between a binary system and a fast-moving perturber, we simulate an ensemble of encounters using the $N$-body simulation code \textsc{rebound}~\cite{Rein:2011uw}. Our configuration consists of an equal-mass binary system with a fixed semimajor axis $a^{\star}$. We take the binary to be circular, i.e., we fix the eccentricity $e$ to zero.  The state of the binary is then determined by four parameters: the cosine of the inclination $\cos\theta_i\in(-1,1)$, the longitude of the ascending node $\Omega\in(0,2\pi)$, the argument of pericenter $\omega\in(0,2\pi)$, and the mean anomaly $M\in(0,2\pi)$ at some fixed time. We randomly sample these parameters while holding the initial conditions of the perturber fixed.

The perturber, with mass $m_3 \ll m_{12}$, is placed at coordinates $(x,y,z) = (b,0,z_0)$. The squared impact parameter $b^2$ is sampled uniformly from the interval $(0, b_{\max}^2)$, where $b_{\max}$ is determined by \cref{eq:impact-parameter} with $r_p$ randomly selected from the interval $[0,a^\star]$. The final result is not sensitive to $z_0$ as long as $z_0 \gg b_{\max}$. We take $z_0=25 b_{\max}$ in our computations. The perturber is given an initial velocity $\bb v_3 = (0, 0, -v_3)$. We evolve the system using the \texttt{ias15} integrator~\cite{Rein_2014}, and we stop integration when the perturber both \textbf{(1)} exits a sphere with a radius twice that of $z_0$ and \textbf{(2)} has positive mechanical energy.

Some integrations, particularly for slower perturbers, can be computationally expensive because the perturber can be captured into metastable orbit, only exiting the system after many periods. To avoid such lengthy simulations, we impose an upper limit on the integration time. If the integration exceeds \num{e4} steps, the specific simulation is discarded. The binding energy of the binary is calculated both before and after the encounter to assess the energy exchange. This process is repeated \num{e5} times for each impact parameter to determine $\langle C \rangle $. Finally, this procedure is carried out for 25 randomly sampled impact parameters.

From these simulations, we extract the distribution of $\delta E$ as a function of the parameters of the encounter. Empirically, we find that the distribution of $\delta E$ is well approximated by Student's $t$ distribution with a scale parameter. We denote this distribution by $\mathcal T(\xi, \nu)$, where $\xi$ is the scale parameter and $\nu$ is the number of degrees of freedom. The pdf is given by
\begin{equation}
    \label{eq:t-pdf}
    f_{\mathcal T}(t) =
    \frac{1}{\xi\sqrt\nu\,\operatorname{B}\left(\frac12,\frac12\nu\right)}
    \left(\frac{\nu}{\nu + (t/\xi)^2}\right)^{\frac12(\nu+1)}
    ,
\end{equation}
where $\operatorname{B}$ denotes the beta function. The (unscaled) $t$ distribution typically arises as the distribution of a random variable $T = \sqrt{\nu}Z/\sqrt{X}$, where $Z\sim\mathcal N(0, 1)$ and $X\sim\chi^2(\nu)$. While it is possible that a similar structure underlies the appearance of the $t$ distribution in this case, the resemblance is only approximate, and it is quite possible that the similarity is purely coincidental. As such, we make no effort to derive this distribution from first principles, but treat it as an empirical fit. (However, see \refcite{2021PhRvX..11c1020G} for an extensive analytical discussion of the energy transfer distribution, with very similar properties to those found here.) When fitting to simulated data, we fit only the $\nu$ parameter, and fix the scale parameter $\xi$ by the requirement that the variance of the fitted distribution matches the variance of the data, using the fact that 
\begin{equation}
    \label{eq:t-var}
    \operatorname{Var}(\mathcal T) = \frac{\nu\xi^2}{\nu - 2}
    .
\end{equation}
Accordingly, $\nu$ is the only free parameter. We do not constrain $\nu$ to be an integer. Note that this is only self-consistent if $\nu > 2$. As $\nu \to 2^+$, the variance diverges, and our analysis assumes that the variance of $\delta E$ is well-defined. For the case of an equal-mass binary with component masses \qty{1}{\msol} and separation \qty{1}{\au}, we find a best-fit value $\nu = 2.29$ across the entire dataset. 

The resulting fits are shown for several perturber velocities in the left panel of \cref{fig:fits-sigmas}, demonstrating broad compatibility of the empirical data with the $t$ distribution. Moreover, the scaling arguments of \cref{sec:multiple-perturbations} are borne out by these numerical experiments. The right panel of \cref{fig:fits-sigmas} shows the dependence of the fitted $\sigma_E$ on impact parameter (top) and perturber velocity (bottom), approximately matching the expectations from \cref{eq:energy-scaling}.

\subsection{Estimated detection probability}
Now that we have determined the distribution of perturbations $\delta E$ for a benchmark scenario, we can use the estimators in \cref{eq:p-tot,eq:p-hat} to study the observability of perturbers under various conditions. In particular, we can now replace $\prob(|\delta E|>E_{\min})$ with the cdf of the appropriate $t$ distribution from the previous subsection, and we can replace $\sigma_E$ with the square root of the variance computed in our simulations. This means that we can now explicitly evaluate $\hat\pobs$ and the other estimators for a given set of parameters to determine the probability of an observable perturbation occurring in a given system.

\begin{figure}
    \includegraphics[width=\linewidth]{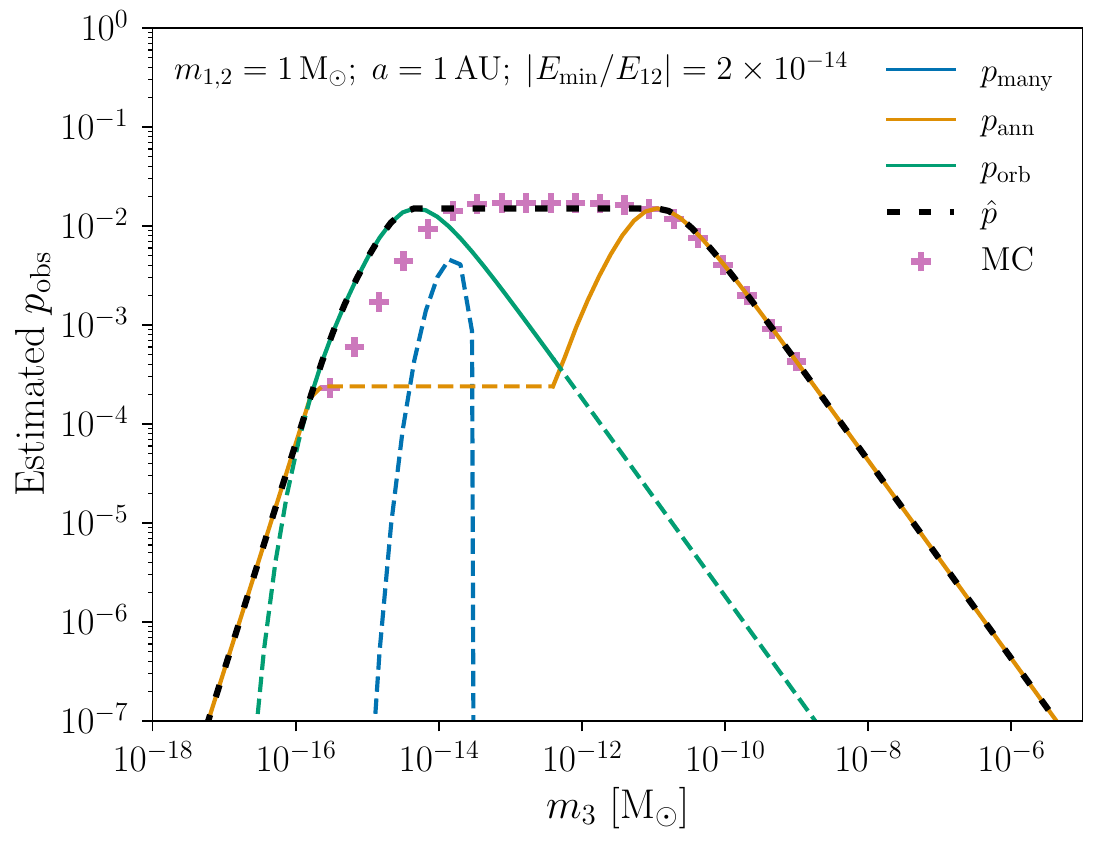}
    \caption{Components of \cref{eq:p-tot,eq:p-hat} for an equal-mass circular binary with component masses $m_1=m_2=\qty{1}{M_\odot}$ and semimajor axis $a=\qty{1}{\au}$. An observing time of \qty{1}{\year} is assumed, with all of the DM in the form of compact objects at mass $m_3$, and the minimal energy transfer for detection is taken to be a \num{2e-14} fraction of the mechanical energy of the binary, chosen to make the curves distinct for illustrative purposes. Blue, orange, and green curves show the values of individual estimators $\pobs_{\many}$, $\pobs_{\ann}$, and $\pobs_{\orb}$, respectively. Each of these curves is solid where it is the greatest of the three, and dashed elsewhere. The dashed black curve shows the value of the estimator $\hat\pobs$. Crosses show results from Monte Carlo sampling of the $t$ distribution (see text).}
    \label{fig:p-estimator}
\end{figure}

First, we consider all four estimators: $\pobs_{\many}$, $\pobs_{\ann}$, $\pobs_{\orb}$, and $\hat\pobs$. \Cref{fig:p-estimator} shows results for each estimator for an equal-mass circular binary system with semimajor axis $a = \qty{1}{\au}$ and equal component masses $m_1=m_2=\qty{1}{\msol}$), varying the perturber mass $m_3$. Here we set $v_3 = \qty{230}{\kilo\meter/\second}$ and $\rho_3= \qty{0.4}{\giga\electronvolt/\centi\meter^3}$, typical of the DM distribution in the Solar neighborhood. In effect, we consider the DM to be made entirely of compact objects of mass $m_3$ for each point. This means that the number density of perturbers scales inversely with $m_3$.

Now, observe that $\pobs_{\many}$ is nonnegligible only for a narrow portion of the mass range. This corresponds to the estimated probability that one would derive from the central limit theorem, with sufficiently many encounters in the observing period. There is a sharp cutoff at high mass as the mean number of encounters drops below $\lambda_{\min}$. On the other hand, there is also a sharp cutoff at low mass: in this regime, there is a large number of encounters whose energy transfers follow the same distribution, and the mean energy transfer falls rapidly, as predicted by \cref{eq:p-obs-regimes}. Indeed, contributions from many similar encounters $(\lambda\gg1)$ are subdominant to the single-encounter contribution $\pobs_{\ann}$, which dominates at both the highest masses and the lowest masses. This means that even in the regime where there are many encounters overall, the probability of a detectable perturbation is still driven by rare single encounters that impart greater energy. In the intermediate regime, where the rate of encounters with $b \lesssim a$ is $\mathcal O(1)$, the interior contribution $\pobs_{\orb}$ dominates. Thus, as anticipated in the definition of $\hat\pobs$, the estimators $\pobs_{\ann}$ and $\pobs_{\orb}$ have distinct peaks. The estimator $\hat\pobs$, shown by the dashed black line in \cref{fig:p-estimator}, bridges the gap between these peaks.

\begin{figure}
    \includegraphics[width=\linewidth]{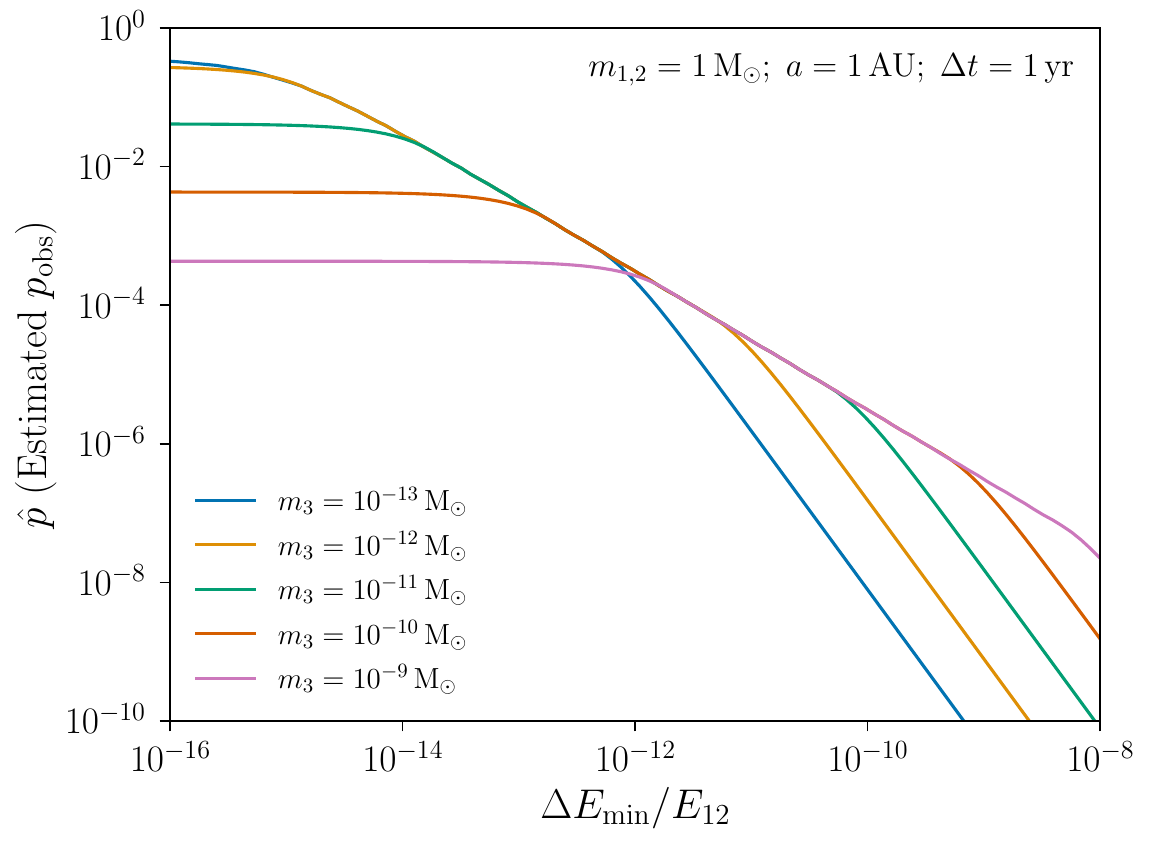}
    \caption{Detection probability as a function of minimum detectable energy for several perturber masses. In the limit of a low detection threshold, the detection probability is limited by the rate of encounters, and increases to 1 at low masses.}
    \label{fig:probability-energy}
\end{figure}

These estimators are meant to approximate the distribution of sums of draws from the distribution of $\delta E$. As such, to the extent that the $t$ distribution is a good approximation of the true distribution of energy transfers, these probabilities can be accurately computed by Monte Carlo methods, i.e., by numerically sampling such multiple draws from the underlying distribution. For each perturber mass $m_3$, we first sample the Poisson-distributed number of encounters during the observing period. For each encounter, we then sample $\delta E$ from the distribution described in \cref{eq:t-pdf,eq:t-var}, and sum these samples to obtain $\Delta E$. We repeat this process many times to obtain the distribution of $\Delta E$ at each $m_3$, and then compute the probability that $\Delta E$ exceeds $\Delta E_{\min}$. The resulting probabilities are shown by the crosses in \cref{fig:p-estimator} (``MC''). Despite the many approximations we have made in this section, the detection probabilities computed by Monte Carlo are in remarkably good agreement with the analytical estimator $\hat\pobs$ (dashed black).

\begin{figure*}\centering
    \includegraphics[width=\linewidth]{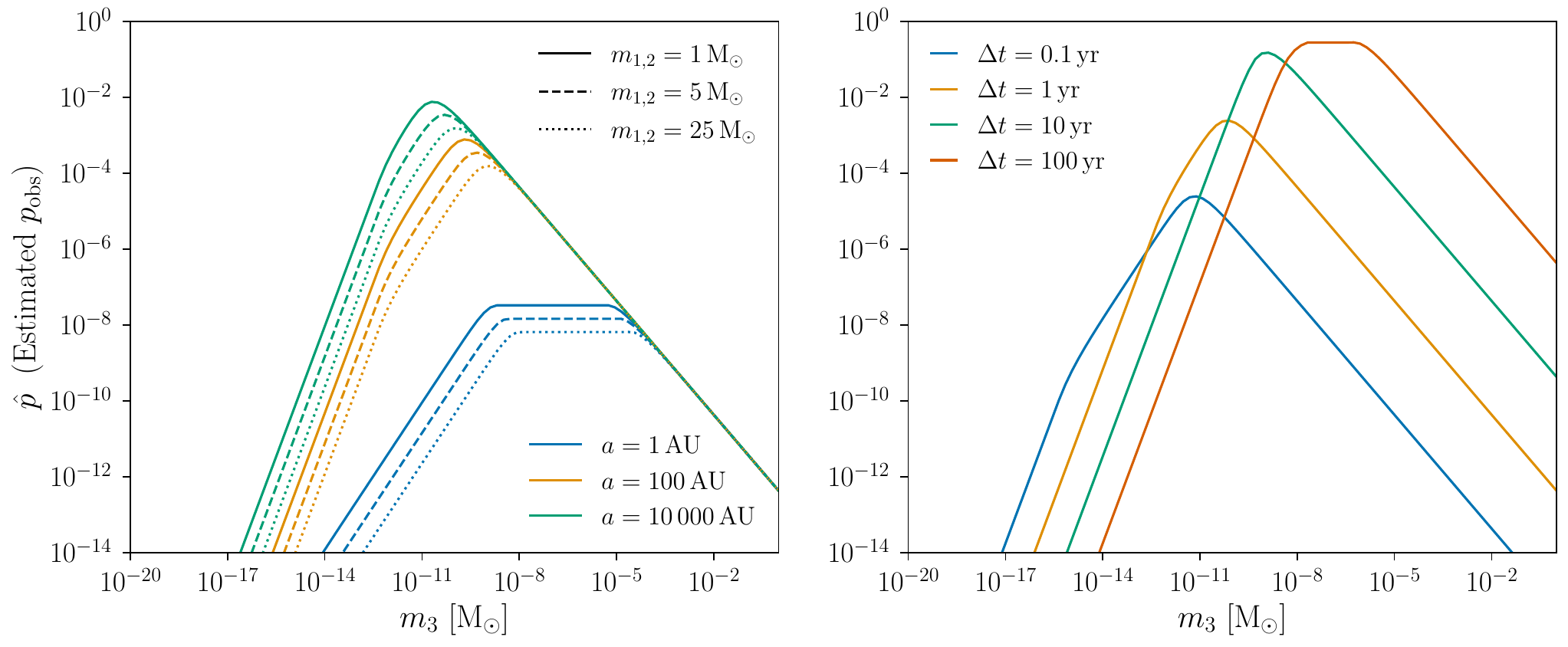}
    \caption{Detection probability estimator $\hat\pobs$ for varying binary parameters and observation times. Both panels assume a minimum detectable energy $\Delta E_{\min}$ corresponding to an average $\dot P_{12} = \num{e-8}$ over the observing period, where $P_{12}$ is the period of the binary. \textit{Left:} varying binary component masses (line styles) and semimajor axes (colors). The observing time is fixed to \qty{1}{\year}. \textit{Right:} varying observation time. The component masses are fixed to $m_{1,2} = \qty{1}{\msol}$, and the semimajor axis is fixed to $\qty{e3}{\au}$.}
    \label{fig:p-scaling}
\end{figure*}

We can now use $\hat\pobs$ to understand the probability of an observable perturbation occuring in various systems. \Cref{fig:probability-energy} shows $\hat\pobs$ as a function of the threshold energy transfer for a detection, $\Delta E_{\min}$, in ratio to the total energy of the binary, $E_{12}$. At high thresholds, larger perturber masses ($m_3$) result in a higher probability of an observable perturbation. But at low thresholds, the detection rate is bottlenecked by the encounter rate, so higher masses, corresponding to lower number densities, are disfavored. Even at very low thresholds, the estimator $\hat\pobs$ only increases to 1 when $m_3$ is low enough to ensure a sufficiently high encounter rate. \Cref{fig:p-scaling} shows the behavior of $\hat\pobs$ as a function of $m_3$ for varying binary masses, semimajor axes, and observation times. For illustration purposes, the detectable energy threshold is set to correspond to a fixed (dimensionless) time derivative of the orbital period $P_{12}$, enforcing that $\dot P_{12} \geq \num{e-8}$. It is immediately obvious from the figure that when fixing precision, wider binaries offer much greater probabilities of detectable perturbations. Binaries with lower component masses are also favored, since transferring the same amount of energy results in a larger change in the period.

The behavior with increasing observation time, shown in the right panel of \cref{fig:p-scaling}, is also easy to understand. We are seeking a stochastic signal, so the probability of a detectable perturbation, $\hat\pobs$, is largest when about one large perturbation can be expected, i.e., when the mean number of perturbers with the appropriate parameters is $\mathcal O(1)$. The mean number of perturbations, $N$, increases linearly with the observing time, and the size of the total perturbation follows the scaling of a random walk, proportional to $\sqrt{N}$. Thus, the energy transfer scales with $\sqrt{\Delta t}$, so the time-averaged hardening rate scales as $1/\sqrt{\Delta t}$. This means that if the precision on the average rate of hardening over the observing time is held fixed, a long observing time sufficient to observe many encounters is \emph{disfavored.} This is visible at low masses (high number densities), where the detection probability decays with observing time. On the other hand, the probability of observing a rare encounter with a heavy perturber, with rate already suppressed by low number density, is enhanced with increased observing time. Thus, the probability curve shifts mainly to the right (higher masses) rather than up (to higher probabilities). This should not be interpreted to suggest that longer observations are inherently less powerful for probing low masses. It is strictly a consequence of fixing the precision on the average hardening \emph{rate}.

It is also apparent from \cref{fig:p-scaling} that with some of the parameter combinations exhibited here, wide binaries would plausibly probe PBH DM in the unconstrained asteroid-mass range. Motived by this observation, we now turn to the consideration of real astrophysical systems that might exhibit the appropriate properties to serve as compact object detectors.

\begin{table*}
    \centering
    \setlength{\tabcolsep}{5pt}
    \begin{tabular}{
        l
        S[table-format=2.3]
        S[table-format=2.3]
        S[table-format=1.4]
        S[table-format=1.4]
        S[table-format=2.2]
        c
        S[table-format=1.4]
        l
    }
        \toprule
        System
            & {$m_1$~[\qty{}{\msol}]}
            & {$m_2$~[\qty{}{\msol}]}
            & {$a_{\mathrm{proj}}$~[\qty{}{\au}]}
            & {$e$}
            & {$\theta_i$~[\qty{}{\degree}]}
            & $\dot P_{12}$
            & {$\left\{|\Delta P_{12}|/P_{12}/m_3\right\}^\star$~[\qty{}{\msol^{-1}}]}
            & Ref. \\
        \midrule
        J1713$-$0747
            & 1.33 & 0.29 & 0.0648 & \num{7.49e-5}
            & 71.69 & $\phantom{-}\num{3.4e-13}$ & 1.2923
            & \cite{Zhu:2018etc} \\
        J1903$+$0327
            & 1.67 & 1.03 & 0.211 & 0.4367
            & 77.47 & \num{-3.3e-11} & 0.8804
            & \cite{Freire:2010tf} \\
        J2016$+$1948
            & 1 & 0.29 & 0.302 & \num{1.48e-3}
            & 58.58 & {--} & 0.8137
            & \cite{Gonzalez:2011kt} \\
        J1740$-$3052
            & 1.4 & 19.5 & 1.52 & 0.5789
            & 53 & $\phantom{-3.}3\times10^{-9\phantom{3}}$ & 0.1271
            & \cite{Madsen:2012rs} \\
        B1259$-$63
            & 2 & 19.8 & 2.60 & 0.8698
            & {$154 \pm 3$}  & $\phantom{-}1.4\times10^{-8\phantom{3}}$
            &  0.0900
            & \cite{Miller-Jones:2018waj, Shannon:2013dpa} \\
        \bottomrule
    \end{tabular}
    \caption{A selection of several observed wide binary pulsars with precisely measured orbital decay rate. Here $a_{\mathrm{proj}}$ denotes the projected semimajor axis; $e$ denotes the eccentricity; $\theta_i$ denotes the inclination angle; and $\dot P_{12}$ gives the measured rate of change of the orbital period of the system. (To our knowledge, a value of $\dot P_{12}$ has not been reported in the literature for J2016$+$1948.) The column $\left\{|\Delta P_{12}|/P_{12}/m_3\right\}^\star$ gives the average value of $|\Delta P_{12}|/P_{12}$ measured in a set of simulated individual encounters, in ratio to the mass of the perturber. The simulations are performed with a light perturber, $m_3 \ll m_{12}$, in the regime where the size of the perturbation to the period is linear in $m_3$.}
    \label{tab:systems}
\end{table*}

\section{Realistic systems and observational prospects}
\label{sec:prospects}
Having developed a flexible estimator of the probability of an observable perturbation, we now consider the astrophysical systems that might lend themselves to a detection of compact objects as DM.

\subsection{General features of sensitive systems}
The first consideration that drives the sensitivity to perturbers is the encounter rate itself. Recall that for a given observing time, $\Delta t$, the maximum perturber distance that we consider in this work is given by $b_{\max} = v_3\,\Delta t$, which bounds the encounter rate above by
\begin{multline}
    \Gamma \lesssim n_3v_3\times\pi b_{\max}^2 \\
    =
        \qty{4}{\per\year} \times
        \left(\frac{m_3}{\qty{e-13}{\msol}}\right)^{-1}
        \left(\frac{\Delta t}{\qty{1}{\year}}\right)
        \left(\frac{v_3}{\qty{230}{\kilo\meter/\second}}\right)^3
    .
\end{multline}
The fact that this rate is $\mathcal O(1)$ over a typical observing time for objects in the asteroid-mass range is the very fact that enables the Solar System probes of \refscite{Tran:2023jci,Bertrand:2023zkl,Thoss:2024vae}. While the same is true in principle here, it is important to note that, as shown in \cref{sec:multiple-perturbations}, the width of the energy transfer distribution---i.e., the figure of merit for stochastic perturbations---declines as $(b/a)^{-2}$ for $b \gtrsim a$. Thus, the width of the distribution is greatest in a region which has cross section ${\sim}a^2$, and the purpose of detection is best served by wider binaries.

On the other hand, the other key consideration is the precision with which a perturbation can be measured. In this work, we consider just one possible signal of a perturbation: a change in the period of a binary system. If a binary is hardened or softened, the orbital period will shift slightly. For an equal-mass circular binary, the period, semimajor axis, and total energy are simply related. The period is given by $P_{12} = \sqrt{2\pi a^3/(2Gm_1)}$, whereas the total energy is given by $E_{12} = \ed{-}Gm_1^2/(4a)$. For small perturbations, therefore, we have
\begin{equation}
    \frac{\Delta P_{12}}{P_{12}} =
        \frac{3}{2}\frac{\Delta E_{12}}{E_{12}}
    .
\end{equation}
The best observational prospects are thus found in systems where the period is long (i.e., wide binaries) and can be measured with high precision. These two desiderata are often in conflict, as we will see later on.

Note that our analysis is based largely on circular binaries. Realistically, typical binaries are somewhat eccentric. While more complex, this may in fact offer a more sensitive probe of compact objects than the semimajor axis alone. Indeed, \refcite{Heggie:1975rcz} showed that the change in eccentricity induced by three-body encounters can be much more sigificant than the change in energy.

For the moment, however, we consider energy transfer alone, and we first consider systems where the period is known with exceptional precision. One set of such systems is provided by the Solar System itself, again providing natural motivation for the work of \refscite{Tran:2023jci,Bertrand:2023zkl,Thoss:2024vae} focusing on Solar System objects. But the Solar System is the not the only setting for precision measurements. A natural alternative is provided by binary pulsars, which are extremely well modeled. In the remainder of this section, we evaluate the prospects for compact object detection via three-body encounters with such systems.

\subsection{Observational prospects in binary pulsars}
Many binary pulsar systems have been detected and characterized at extremely high precision, especially as components of pulsar timing arrays for gravitational wave detection~\cite{Hobbs:2009yy,Manchester:2012za,McLaughlin:2013ira,Kramer:2013kea}. The result is a large set of systems with separations comparable to Solar System orbits and, at the same time, reasonably high precision on the measurement of the period. As with the Solar System, such systems have been famously used for many tests of general relativity~\cite{Taylor:1982zz,1986ARA&A..24..537B,Taylor:1992kea,Taylor:1994zz,Stairs:2003eg,Kramer:2006nb,Weisberg:2010zz,Yunes:2013dva}, with extraordinary sensitivity to anomalous accelerations, and it is thus natural to consider using them as a detector for compact objects. In fact, in general terms, several authors have previously considered the possibility that perturbers might leave a mark in pulsar timing data used for gravitational wave detection~\cite{Siegel:2007fz,Seto:2007kj,Clark:2015sha,Kashiyama:2018gsh,Jennings:2019qqz,Dror:2019twh,Ramani:2020hdo}. They have even been used as settings for the study of dynamical friction~\cite{Caputo:2017zqh,Pani:2015qhr}. We now consider the possible use of pulsars for compact object detection in our framework. Note that we only assume sensitivity to changes in the average period over the entire observing window, not transient fluctuations.

\begin{figure*}\centering
    \includegraphics[width=\linewidth]{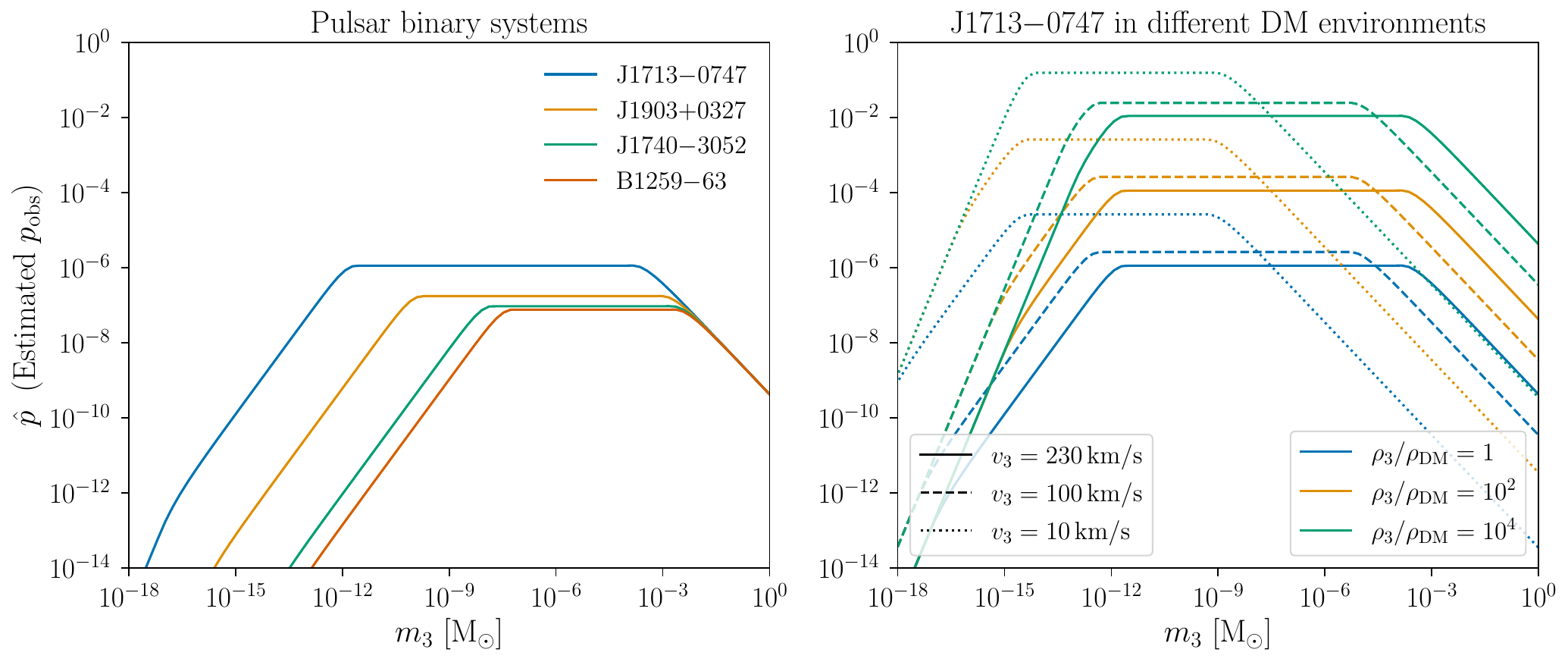}
    \caption{Detection probability estimator $\hat\pobs$ for a variety of astrophysical systems with parameters based on those listed in \cref{tab:systems}. Here, each binary is taken to be circular, with equal component masses given by the mean of $m_1$ and $m_2$. The minimum detectable perturbation to the period is conservatively taken to be equal to the measured value of $\Delta P_{12}$, i.e., $\dot P_{12}$ integrated over the observing time. Since $\dot P_{12}$ is not available for J2016$+$1948, we exclude this system from consideration here.}
    \label{fig:astrophysical}
\end{figure*}

The properties of a set of example binary pulsars is given in \cref{tab:systems}, including the observed rate of change of the period, $\dot P_{12}$. For each of these systems, we conduct an ensemble of simulations of three-body encounters with a light perturber of mass $m_3^\star \ll \min\{m_1, m_2\}$, and we determine the average magnitude of the change in the period, $|\Delta P_{12}|$, resulting from each encounter. Since these simulations are in the test-particle regime, with $m_3^\star \ll m_{1,2}$, the energy imparted, and thus the change to the period, is linear in the mass of the perturber. Accordingly, $|\Delta P_{12}|$ can be predicted for perturbers of smaller masses by a simple rescaling. We report the results of these simulations in \cref{tab:systems} in the form $|\Delta P_{12}|/P_{12}/m_3^\star$, so the expected value of $|\Delta P_{12}|/P_{12}$ for a single encounter can be obtained by multiplying this column by the perturber mass. (Note that while we report the results in units of \qty{}{\msol^{-1}}, linearity only holds for $m_3 \ll \qty{}{\msol}$.)

To estimate the sensitivity of these systems to perturbers of varying masses, we evaluate our estimator $\hat\pobs$ under a set of approximations. We consider a \qty{10}{\year} observation of each of these systems, and approximate the rate of perturbations by taking each system to be an equal-mass circular binary, holding the total mass and semimajor axis fixed. We conservatively estimate the minimum detectable perturbation by taking the precision on $\Delta P_{12}$ to be equal to the measured value of $\Delta P_{12}$ implied by $\dot P_{12}$ over the entire oberving period. (We exclude J2016$+$1948, since $\dot P_{12}$ is not available for this system.) We show the resulting estimated probability of detecting a perturbation for each of these systems in the left panel of \cref{fig:astrophysical}. These probabilities are generally small, and are dominated by the system J1713$-$0747, which offers a compromise between semimajor axis and precision. In the right panel of \cref{fig:astrophysical}, we thus show the probabilities that would be obtained in the case of DM overdensities (i.e., large $\rho_3$) or cold features (i.e., small $v_3$). J1713$-$0747 has percent-level probability of an observable perturbation in the case of a \num{e4} overdensity of DM, increasing to $\mathcal O(10\%)$ in a cold feature with $v_3\sim\qty{10}{\kilo\meter/\second}$.

All of these probabilities are still well below 1. This is not surprising: it is well known that no single binary pulsar system can constrain compact objects as DM. But the possibility of a stochastic effect offers the prospect for a new class of search. Rather than studying a single system, it is plausible that a combination of several or many systems with non-negligible $\hat\pobs$ might yield a nontrivial constraint. Our estimator provides a quick method of estimating the sensitivity enabled by observations of various systems, or combinations thereof.

\begin{figure}
    \includegraphics[width=\linewidth]{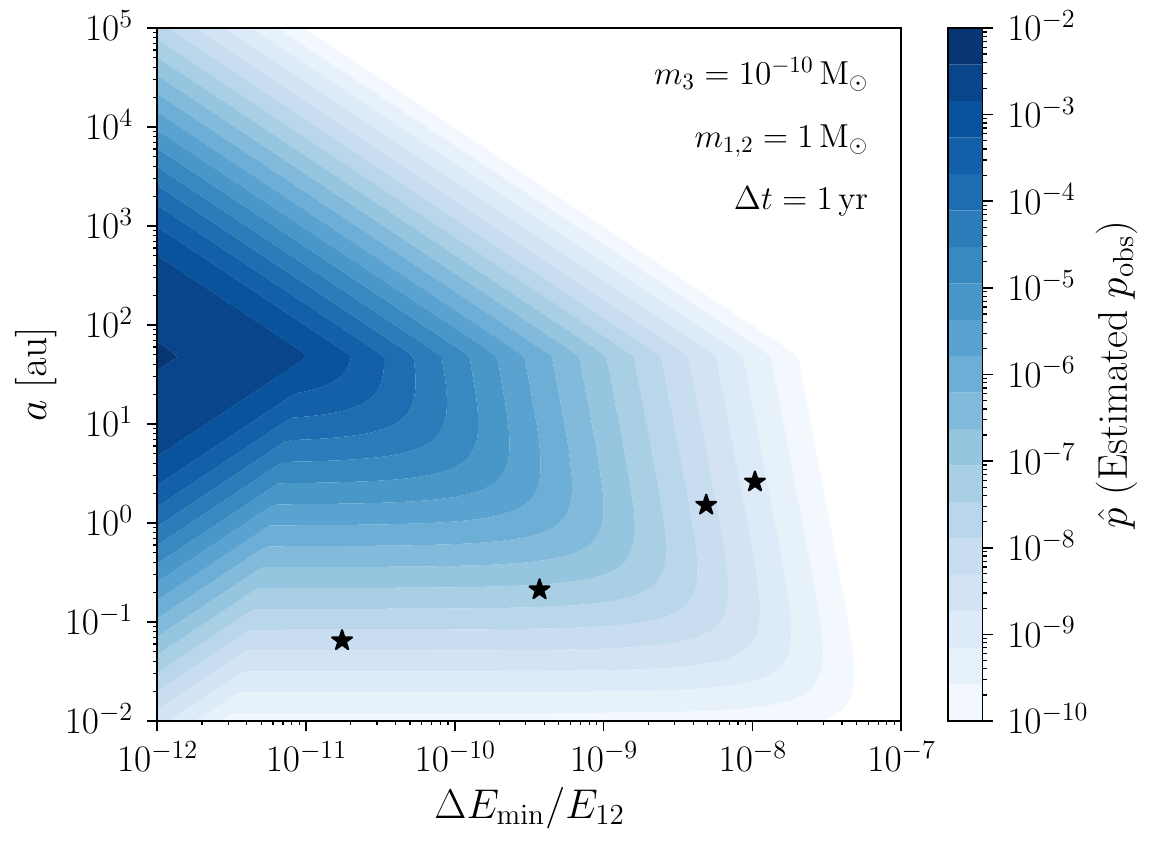}
    \caption{Detection probability as a function of minimum detectable energy and binary semimajor axis for a fixed perturber mass $m_3 = \qty{e-10}{\msol}$ and binary component masses $m_1=m_2=\qty{1}{\msol}$, over an observing period of \qty{1}{\year}. The binary pulsar systems of \cref{tab:systems} are marked with stars.}
    \label{fig:separation-threshold}
\end{figure}

For example, note that at fixed $m_3$ and $\Delta t$, under the assumptions made here, there is an optimal separation for a system, beyond which sensitivity is degraded. This is immediately visible in \cref{fig:separation-threshold}, which illustrates the relative importance of semimajor axis and observational precision. For the case shown here, with perturber mass $m_3 = \qty{e-10}{\msol}$, binary component masses $m_{1,2} = \qty{1}{\msol}$, and an observing time of $\Delta t = \qty{1}{\year}$, the optimal separation is \num{10}--\qty{100}{\au}. This is considerably larger than the typical separations of binary pulsars. \Cref{fig:separation-threshold} also demonstrates the tradeoff between precision and separation: the binary pulsars of \cref{tab:systems} are marked by stars, and the precision of measurement is degraded at higher separations. Thus, robust constraints on compact objects may need to wait for higher precisions or a new set of systems. For pulsars in particular, the discovery of a vast number of new systems is anticipated at the Square Kilometer Array (SKA), which is also expected to considerably improve the precision of pulsar timing measurements~\cite{Keane:2014vja,Shao:2014wja}. The associated improvement in precision on pulsar binary parameters should be significant, but a quantitative estimate is nontrivial, and beyond the scope of the present work. We note that several new types of system will be accessible in the coming decades with future gravitational wave experiments, which also offer precise measurements of orbital periods. We defer detailed consideration of these systems and their potential sensitivities to future work.

\section{Conclusions}
\label{sec:conclusions}
The purpose of this work is twofold. On the one hand, we advance a novel probe of dark compact objects such as PBHs, with potential applications to a challenging mass range. On the other hand, as a purely astrophysical problem, we complete the panoply of three-body encounter outcomes shown in \cref{fig:schematic}, identifying rich structure in a portion of the parameter space often regarded as empty in the past.

Beginning with the latter component, we have shown in this work that dynamical effects from three-body encounters can in fact have a hardening or softening effect, despite astrophysical lore that their speed renders this impossible. While the regime of fast perturbers faces a suppression in energy transfers compared to slow perturbers, this is compensated for by stochasticity: in the limit of few perturbers, the net effect is potentially discernible. This provides a satisfying coda to the story originally started by the work of \refscite{Heggie:1975rcz,Gould:1990bk,Quinlan:1996vp}: the average shrinks to zero, but the width does not, leaving a stochastic effect in this regime. Thus, the analogue of the hardening rate is the rate of detectable perturbations in either direction. Here, we have developed a semianalytical framework for computing this rate, summarized in our statistical estimator $\hat\pobs$.

It is easy to understand why such effects have not been considered in the past. Generally, one is interested in the long-term evolution of systems rather than perturbations of the kind we study here. Over a long timescale, the effects we identify in this work vanish. In cases such as supermassive black hole evolution, three-body hardening is only relevant at stages where the velocity of the binary is large enough compared to the Galactic halo dispersion. At other times, dynamical friction is the correct framework, and exhibits different behavior.

Our sensitivity projection also explains the utility of objects in the Solar System as probes for dark compact objects, as discussed in detail by \refscite{Tran:2023jci,Thoss:2024vae}. The projections in these works assume resolution of the orbit of e.g. Mars that corresponds to a precision of $\mathcal O(\num{e-12}\textnormal{--}\num{e-14})$ on the period. Under these conditions, the plateau in the detection probability is in the range $(\hat m_\orb, \hat m_\ann) \simeq (\num{e-14}, \num{e-8})\qty{}{M_\odot}$, and within this plateau, the detection probability is $\mathcal O(1)$, with $\hat\pobs \simeq 30\%$. Thus, our framework reproduces the statement that the Solar System can plausibly detect compact objects in the asteroid-mass window. Moreover, we can now describe Solar-System--based detection proposals in the same framework as three-body hardening (at low masses and velocities) and three-body softening (at high masses and velocities).

Pragmatically, the utility of the method we develop here is that it can be readily applied to potential targets well beyond the Solar System. The challenge is to identify systems with the optimal balance of large separation and high-precision period measurements. In this work, we have applied our computation to binary pulsars, which have their periods measured to high precision. The size of perturbations is sensitive to the velocity of the perturbers, making this technique a uniquely powerful probe of DM structures with low velocity dispersion. We verify that the currently observed wide binary pulsars cannot constrain the Galactic halo in the absence of an overdensity or cold feature, but future measurements can potentially improve the precision to enable sensitivity to the diffuse Galactic halo.

In searches involving wide binary systems, current precision measurements primarily rely on astrometric methods, such as those from Gaia observations~\cite{El-Badry:2024vjt}, which have been used to constrain massive compact objects~\cite{Ramirez:2022mys}. Due to Gaia's relatively short observation baseline, of order 10 years, it cannot directly measure orbital decay rates over multiple successive orbits for binaries with very wide separations. For example, the semimajor axis of a binary with \qty{1}{M_\odot} components with an orbital period of 10 years is about \qty{6}{\au}, so for these masses, multiple orbits can only be observed for systems with considerably smaller separations. This does not preclude measuring orbital decay rates with sufficiently high precision on partial orbits, and the next-generation astrometry mission, Theia, is expected to enable improved measurements with its higher astrometric accuracy and longer observation time~\cite{Theia:2017xtk}. Future prospects include binaries that can be observed and monitored precisely with the Laser Interferometer Space Antenna (LISA)~\cite{Kupfer:2018jee,Lamberts:2019nyk}. The double white dwarf binaries expected to measured by LISA would exhibit baseline orbital decay rates as small as $\mathcal{O}(\qty{e-16}{\per\second})$ due to gravitational wave emission. However, the separations of such binaries remain below \qty{1}{\au}, facing challenges similar to those encountered with binary pulsars.

Our results also clarify the connection between the regime in which multiple three-body encounters are relevant and that in which dynamical friction is relevant. In the limit of a large number of encounters with light perturbers, the rate of hardening from three-body encounters regresses to its average value, and vanishes, leaving only dynamical friction. On the other hand, when the number of close encounters over the observing period is small, the individual perturbations are substantial, and do not efficiently average to zero. (Interestingly, ultralight DM may be detectable by the same means, since stochastic fluctuations in such a field can similarly perturb binaries over short timescales~\cite{Foster:2025csl,Foster:2025nzf}.)

Note that while pulsar binary systems that we consider in \cref{tab:systems} are sensitive to DM only in the presence of overdensities, such overdensities are also sufficient to observe the effects of particle DM via dynamical friction in the same systems, as discussed in detail by \refscite{Pani:2015qhr,Caputo:2017zqh}. Thus, systems sensitive enough to detect dynamical friction are likely also sensitive to stochastic three-body interactions. This suggests that detection of DM by these means would also readily reveal whether the DM is particle-like or compact-object--like. In some regimes, dynamical friction and stochastic perturbations appear quite differently in timeseries data, with the former entering as a smooth effect and the latter as a set of transient events. Even in the absence of such timeseries data, since dynamical friction and three-body perturbations scale differently with the properties of the binary, comparing the size of the effect over a long time in two or more systems can provide robust discrimination between these two scenarios.

Finally, we stress that in this work, we have focused on the stochastic effects that dominate when the perturbers are fast relative to the binary components. While this is typically the case for realistic binaries in the Galactic halo, there are exceptions, e.g. in short-period systems or in very cold DM environments. Thus, there are some scenarios where traditional time-averaged three-body hardening may apply, and constraints on light perturbers may be derived from population properties of hard binaries. We defer consideration of these systems to future work.

Our results point the way towards a powerful method to constrain the mass of individual DM constituents even in the challenging asteroid-mass window. Few-body dynamical probes of the kind we propose here may have provided the first-ever bound on the DM mass~\cite{1969Natur.224..891V}. Remarkably, as we have shown in this work, there is yet further information to be extracted from these processes---and they may yet provide decisive bounds on compact objects as DM.

\begin{acknowledgments}
We thank Alan Guth, Yacine Ali-Ha\"imoud, David I. Kaiser, Annika H. G. Peter, and Jaco de Swart for valuable conversations.
The work of BB is supported by the Avenir Fellowship.
The work of BVL is supported by the MIT Pappalardo Fellowship.
The research activities of KS and TX are supported in part by the U.S. National Science Foundation under Award No. PHY-2412671. BVL and KS would like to thank the Aspen Center for Physics, which is supported by National Science Foundation grant PHY-2210452, for hospitality during the course of this work. KS and TX also wish to acknowledge the Center for Theoretical Underground Physics and Related Areas (CETUP*) and the Institute for Underground Science at SURF for hospitality and for providing a stimulating environment.
We extend our sincere gratitude to the OU Supercomputing Center for Education and Research (OSCER) for their state-of-the-art resources that were instrumental in enabling us to conduct extensive simulations.
\end{acknowledgments}

\bibliography{references}

\end{document}